\documentclass[apjl,useAMS,usenatbib]{emulateapj}
\usepackage{times}
\usepackage{natbib}
\usepackage{apjfonts}
\usepackage{amssymb}
\usepackage{epsfig}

\def\hi{H{\sc i}}
\def\nhi{\rm N_{HI}}

\def\mgtwo{Mg{\sc ii}$\lambda$2796}
\def\wmgtwo{W_0^{\lambda 2796}}
\newcommand{\kms}{km~s$^{-1}$}
\newcommand{\cm}{cm$^{-2}$}

\newcommand{\ts}{T_s}
\newcommand{\beq}{\begin{equation}}
\newcommand{\eeq}{\end{equation}}

\shorttitle{A metallicity-spin temperature relation in DLAs}
\shortauthors{Kanekar et al. }

\begin{document}

\title{A metallicity-spin temperature relation in damped Lyman-$\alpha$ systems}


\author{Nissim Kanekar\altaffilmark{1},
Alain Smette\altaffilmark{2},
Frank H. Briggs\altaffilmark{3},
Jayaram N. Chengalur\altaffilmark{4}
}

\altaffiltext{1}{National Radio Astronomy Observatory, 1003 Lopezville Road, Socorro, NM87801, USA; nkanekar@nrao.edu}
\altaffiltext{2}{European Southern Observatory, Alonso de Cordova 3107, Casilla 19001, Vitacura, Santiago, Chile}
\altaffiltext{3}{RSAA, The Australian National University, Mount Stromlo Observatory, ACT 2611, Australia}
\altaffiltext{4}{National Centre for Radio Astrophysics, TIFR, Ganeshkhind, Pune-411007, India}

\begin{abstract}
We report evidence for an anti-correlation between spin temperature $T_s$ and metallicity [Z/H],
detected at $3.6 \sigma$ significance in a sample of 26~damped Lyman-$\alpha$ absorbers (DLAs)
at redshifts $0.09 < z < 3.45$. The anti-correlation is detected at $3 \sigma$ significance 
in a sub-sample of 20~DLAs with measured covering factors, implying that it does not stem 
from low covering factors. We obtain $T_s = (-0.68 \pm 0.17) \times {\rm [Z/H]} + (2.13 \pm 0.21)$ 
from a linear regression analysis. Our results indicate that the high $T_s$ values found 
in DLAs do not arise from differences between the optical and radio sightlines, but are 
likely to reflect the underlying gas temperature distribution. The trend between $T_s$ and [Z/H] 
can be explained by the larger number of radiation pathways for gas cooling in galaxies with 
high metal abundances, resulting in a high cold gas fraction, and hence, a low spin temperature. 
Conversely, low-metallicity galaxies have fewer cooling routes, yielding a larger warm gas 
fraction and a high $T_s$. Most DLAs at $z>1.7$ have low metallicities, 
[Z/H]~$< -1$, implying that the H{\sc i} in high-$z$ DLAs is predominantly warm. The 
anti-correlation between $T_s$ and [Z/H] is consistent with the presence of a mass-metallicity 
relation in DLAs, suggested by the tight correlation between DLA metallicity and the 
kinematic widths of metal lines. Most high-$z$ DLAs are likely to arise in galaxies with 
low masses ($M_{\rm vir} < 10^{10.5} M_\odot$), low metallicities ([Z/H]$< -1$, and 
low cold gas fractions.
\end{abstract}


\section{Introduction}

Damped Lyman-$\alpha$ systems, with \hi\ column densities $\nhi \ge 2 \times 10^{20}$~\cm,
are the high-$z$ gas-rich counterparts of today's normal galaxies, and crucial to 
understanding galaxy evolution. However, despite much observational effort, our
knowledge of the typical size, structure and internal conditions of high $z$ DLAs is yet 
limited (e.g. \citealp{wolfe05}). A controversial issue is the \hi\ temperature 
distribution in the absorbers, whether most of the \hi\ is in a warm, low-density 
phase (the warm neutral medium, WNM, with kinetic temperature $T_k \sim 5000 - 8000$~K; 
\citealt{wolfire95}), or whether a significant fraction is in a high density, cold phase 
(the CNM, with $T_k \sim 40 -200$~K). For DLAs towards radio-loud quasars, this can 
be determined by combining the optical depth in the redshifted \hi~21cm line with the 
\hi\ column density measured from the Lyman-$\alpha$ profile to obtain the 
column-density-weighted harmonic mean spin temperature $\ts$ along the sightline 
(e.g. \citealp{kanekar04}). High-$z$ DLAs have been shown to have higher $\ts$ values 
($\gtrsim 700$~K; e.g. \citealt{wolfe79,carilli96,kanekar03}) than those typically 
found in the Milky Way or local spirals ($\ts \lesssim 300$~K; \citealt{braun92}).
The simplest interpretation of this, in the context of stable two-phase models, 
is that high-$z$ DLAs contain larger WNM fractions than local disks 
\citep{carilli96,chengalur00}.  Conversely, \citet{wolfe03b} argue that the 
detection of CII* absorption in a number of high-$z$ DLAs indicates sizeable 
CNM fractions ($\sim 50$\%) in half the DLA population.

A plausible cause for the putative higher WNM fractions in high-$z$ DLAs is their 
typically-low metallicity [Z/H], implying fewer routes for gas cooling. If high $\ts$ 
values in DLAs arise due to low absorber metallicities, one would expect an 
anti-correlation between $\ts$ and [Z/H], with low-$\ts$ DLAs having high metallicities, 
and vice-versa \citep{kanekar01a}. We report here the detection of the predicted 
anti-correlation between $\ts$ and [Z/H], supporting the conclusion that \hi\ in 
high-$z$ DLAs is predominantly in the warm neutral medium.

\section{The sample}
\label{sec:sample}

\setcounter{table}{0}
\begin{table*}
\begin{center}
\caption{The full sample of 26 DLAs.
\label{tab:met}}
\begin{tabular}{|c|c|c|c|c|c|c|c|c|}
\hline
QSO   & $z_{\rm abs}$ & N$_{\rm HI}$         & $f$ & $\ts$        & [Z/H]            & [Z/Fe]          & Z,Fe  &  Ref. \\
      &               & $\times 10^{20}$ \cm &     &  K    	  &                  &                 &       &       \\
\hline
      &               &                 &          &              &                  &                 &       &       \\
0738+313  &  0.0912   &  $15 \pm 2    $ & $0.98$ & $775 \pm 100 $ & $<-1.14        $ & $< 0.48$        & Zn,Fe & 1 \\
0738+313  &  0.2212   &  $7.9 \pm 1.4 $ & $0.98$ & $870 \pm 160 $ & $<-0.7 	   $ & $< 0.74$        & Zn,Cr & 1 \\
0952+179  &  0.2378   &  $21.0 \pm 2.5$ & $0.66$ & $6470 \pm 965$ & $<-1.02        $ & $< 0.63$        & Zn,Cr & 1 \\
1127$-$145&  0.3127   &  $51 \pm 9    $ & $0.9 $ & $820 \pm 145 $ & $-0.90 \pm 0.11$ & $-$             & Zn,$-$& 2 \\
1229$-$021&  0.3950   &  $5.6 \pm 0.9 $ & $0.42$ & $95 \pm 15   $ & $-0.45 \pm 0.14$ & $> 0.83$        & Zn,Fe & 3 \\
0235+164  &  0.5242   &  $50 \pm 10   $ & $1.0 $ & $210 \pm 45  $ & $-0.14 \pm 0.17$ & $1.73 \pm 0.24$ & X-ray,Fe & 4 \\
0827+243  &  0.5247   &  $2.0 \pm 0.2 $ & $0.7 $ & $330 \pm 65  $ & $-0.62 \pm 0.05$ & $-$             & Fe+0.4,$-$ & 1 \\
1122$-$168&  0.6819   &  $2.8 \pm 1   $ & $-$    & $>1445 $       & $<-1.47        $ & $< -0.15$       & Zn,Fe & 1,2 \\
1331+305  &  0.6922   &  $17.8 \pm 0.8$ & $0.9 $ & $965 \pm 105 $ & $-1.35 \pm 0.03$ & $0.28 \pm 0.03$ & Zn,Fe & 5 \\
0454+039  &  0.8596   &  $4.9 \pm 0.2 $ & $0.5 $ & $>690 $        & $-0.99 \pm 0.12$ & $0.00 \pm 0.14$ & Zn,Fe & 1 \\
2149+212  &  0.9115   &  $5 \pm 1     $ & $-$    & $>2700 $       & $<-0.93        $ & $-$             & Zn,$-$& 6 \\
1331+170  &  1.7764   &  $15 \pm 1.4  $ & $0.72$ & $1260 \pm 335$ & $-1.20 \pm 0.04$ & $0.847\pm0.008$ & Zn,Fe & 7 \\ 
1157+014  &  1.9436   &  $63 \pm 15   $ & $0.63$ & $1015 \pm 255$ & $-1.40 \pm 0.10$ & $0.38 \pm 0.16$ & Zn,Fe & 8 \\ 
0458$-$020&  2.0395   &  $60 \pm 10   $ & $1.0 $ & $560 \pm 95  $ & $-1.27 \pm 0.07$ & $0.47 \pm 0.05$ & Zn,Fe & 7 \\
0311+430  &  2.2898   &  $2.0 \pm 0.5 $ & $-$    & $140 \pm 35  $ & $>-0.6         $ & $-$             & Si,Fe & 9 \\
0432$-$440&  2.3021   &  $6.0 \pm 1.4 $ & $-$    & $>995 $        & $-1.12 \pm 0.15$ & $0.33 \pm 0.21$ & Si,Fe & 10 \\
0438$-$436&  2.3474   &  $6.0 \pm 1.4 $ & $0.59$ & $900 \pm 250 $ & $-0.68 \pm 0.15$ & $0.62 \pm 0.21$ & Zn,Fe & 10 \\
0405$-$331&  2.5693   &  $4.0 \pm 0.9 $ & $0.44$ & $>785 $        & $-1.40 \pm 0.15$ & $0.34 \pm 0.21$ & Si,Fe & 10 \\
0913+003  &  2.7434   &  $5.5 \pm 1.3 $ & $-$    & $>800 $        & $-1.47 \pm 0.15$ & $0.12 \pm 0.21$ & Si,Fe & 10 \\
1354$-$170&  2.7799   &  $2.0 \pm 0.7 $ & $-$    & $>795 $        & $-1.86 \pm 0.16$ & $0.67 \pm 0.10$ & Si,Fe & 7 \\
2342+342  &  2.9084   &  $13 \pm 3    $ & $0.71$ & $>1705 $       & $-1.23 \pm 0.15$ & $0.36 \pm 0.13$ & Zn,Fe & 11 \\
0537$-$286&  2.9742   &  $2.0 \pm 0.5$  & $0.47$ & $> 520$        & $< -0.44$        & $-$             & Zn,Fe & 10 \\
0336$-$017&  3.0619   &  $15.8 \pm 3.6$ & $0.68$ & $>6670  $      & $-1.40 \pm 0.10$ & $0.37 \pm 0.03$ & S,Fe  & 7 \\
0335$-$122&  3.1799   &  $6.0 \pm 1.4$  & $0.62$ & $>1850  $      & $-2.56 \pm 0.15$ & $0.05 \pm 0.21$ & Si,Fe & 10 \\
0201+113  &  3.3875   &  $18 \pm 3   $  & $0.76$ & $1050 \pm 175$ & $-1.26 \pm 0.15$ & $0.18 \pm 0.21$ & S,Fe  & 12 \\
1418$-$064&  3.4482   &  $2.5 \pm 0.6$  & $0.69$ & $>680   $      & $-1.48 \pm 0.15$ & $0.24 \pm 0.21$ & Si,Fe & 10 \\
      &               &                 &        &                &                  &                 &       &   \\
\hline
\end{tabular}
\vskip -0.01in
Notes: The $\ts$ values have been re-computed uniformly by Kanekar et al. (\textit{in prep.}).
References for [Z/H] values are mostly to literature compilations, which contain the 
original references; all values are scaled to the solar abundances of \citet{lodders03}. 
(1)~\citet{kulkarni05}; (2)~Kanekar et al. (\textit{in prep.}); 
(3)~\citet{boisse98}; (4)~\citet{junkkarinen04}; (5)~\citet{wolfe08}; (6)~\citet{nestor08}; 
(7)~\citet{prochaska07}; (8)~\citet{ledoux06}; (9)~\citet{ellison08}; (10)~\citet{akerman05};
(11)~\citet{prochaska03b}; (12)~\citet{ellison01b}.
\end{center}
\end{table*}

Over the last decade, we have carried out \hi~21cm absorption studies of DLAs 
towards compact, radio-loud quasars to measure their spin temperatures (e.g. 
\citealt{kanekar03,kanekar06,kanekar07,york07}), and have also measured 
the DLA covering factors through low-frequency very long baseline interferometry 
(VLBI) studies \citep{kanekar09a}. The VLBI images yield the fraction of compact 
radio emission, and thus a lower limit to the DLA covering factor. However, 
no additional radio emission is detected up to scales of $\sim 1''$, indicating that 
the remaining emission arises from much larger scales ($\gtrsim 10$~kpc), and is 
unlikely to be covered by the foreground DLA. The radio core fraction thus provides 
a good estimate of the DLA covering factor. 

We have also obtained metallicity estimates for most of the DLAs with 
\hi~21cm studies from our own observations or the literature. There are 26~DLAs, 
at $0.09 \lesssim z \lesssim 3.45$, with estimates of both $\ts$ and [Z/H], 
of which 20 have covering factor estimates from low-frequency VLBI studies, and
21~have estimates of dust depletion, [Z/Fe]. The 26~absorbers of the sample are 
listed in Table~\ref{tab:met}, whose columns contain (1)~the quasar name, (2)~the 
DLA redshift, (3)~the \hi\ column density measured from the Lyman-$\alpha$ profile, 
(4)~the DLA covering factor $f$, (5)~the spin temperature $\ts$, or, for \hi~21cm 
non-detections, the $3\sigma$ lower limit to $\ts$ (taking into account the DLA 
covering factor, when known), (6)~the metallicity [Z/H], (7)~the dust depletion 
factor [Z/Fe], (8)~the transitions used for the [Z/H] and [Z/Fe] estimates, and 
(9)~references for [Z/H] and [Zn/Fe] values. In all but two cases, ${\rm Z} \equiv 
{\rm Zn, S, Si}$, in order of preference.  
The exceptions are the $z \sim 0.524$ DLAs towards 0235+164 and 0827+243, where
the metallicities are, respectively, from an X-ray spectrum
\citep{junkkarinen04} and [Z/H]$=$[Fe/H]$+0.4$ [following \citet{prochaska03a}].
The sample contains 14~$\ts$ measurements and 12~lower limits, and 
20~[Z/H] measurements, five upper limits, and one lower limit (towards 0311+430; 
\citealt{ellison08}). Detailed references for the metallicities and spin temperatures 
are provided in Kanekar et al. (\textit{in prep.}).

Note that the sample does not include four systems where the radio emission is clearly 
extended on scales $>> 1''$, and the covering factor is likely to be low, $f << 1$; these are 
the \hi~21cm absorbers at $z \sim 0.437$ towards 3C196 \citep{briggs01} and $z \sim 0.656$ 
towards 3C336 \citep{curran07a} (where the \hi~21cm absorption arises towards extended 
lobes, with little radio flux density associated with the quasar core), and the 
\hi~21cm non-detections at $z \sim 1.3911$ towards QSO~0957+561 and $z \sim 1.4205$ towards 
PKS~1354+258 \citep{kanekar03}. We have also excluded ``associated'' DLAs, lying within 
$\sim 3000$~\kms\ of the quasar, as conditions in these absorbers could be affected
by their proximity to an active galactic nucleus.

Preliminary results of this study showing the anti-correlation between $\ts$ and [Z/H] 
were presented in \citet{kanekar05b}. \citet{curran07c} later also found weak 
evidence for the anti-correlation, but did not have covering factor estimates for
most high-$z$ DLAs, and hence could not rule out the effects of unknown covering factors.
The results of the present {\it Letter} are based on new \hi~21cm data on $11$~absorbers, 
and VLBI estimates of the DLA covering factor for most systems, and yield the first clear 
evidence for a relation between the metallicity and the \hi\ temperature distribution 
in the absorbers.

\begin{figure*}[t!]
\epsfig{file=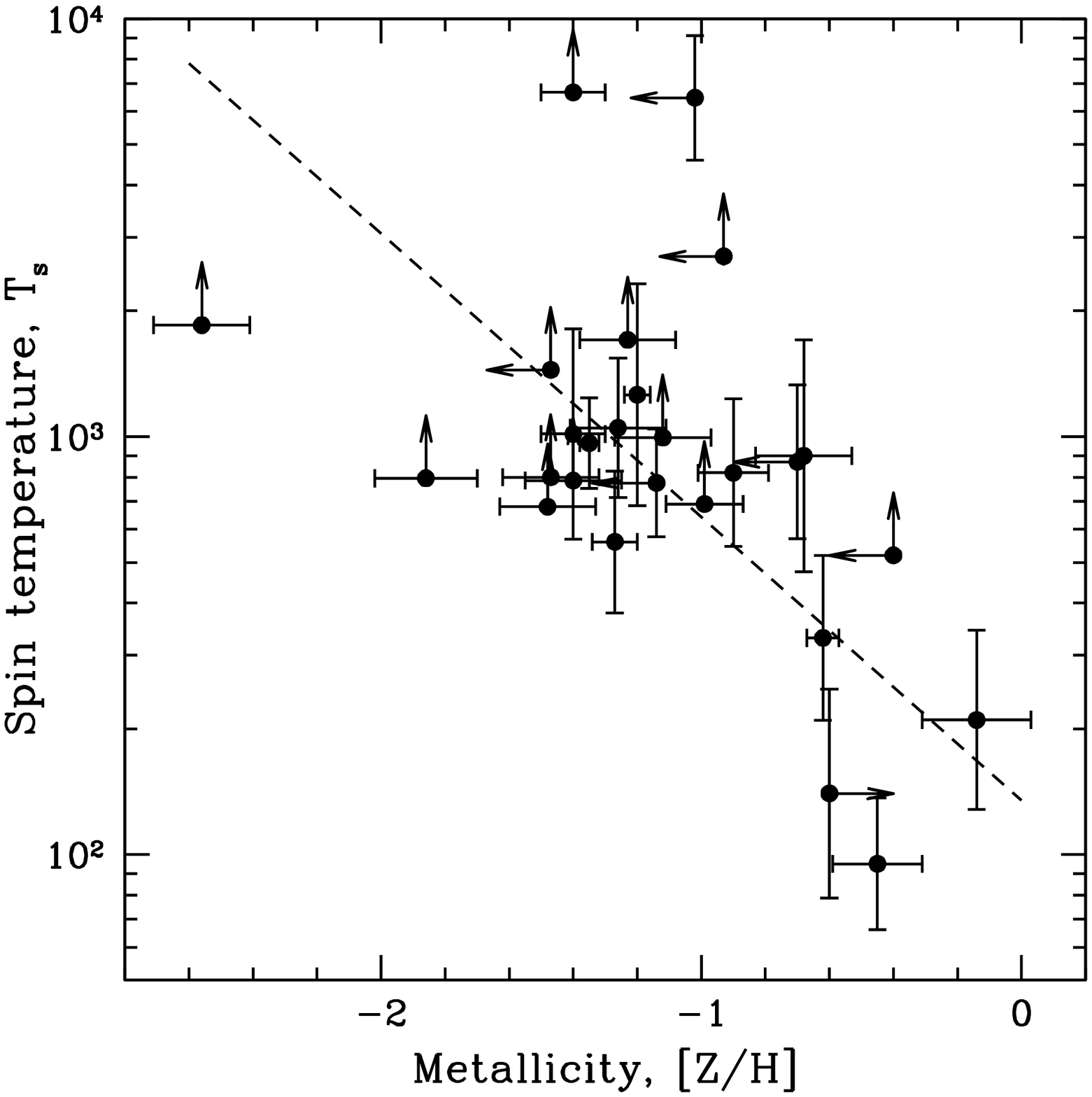,height=3.5truein,width=3.5truein}
\epsfig{file=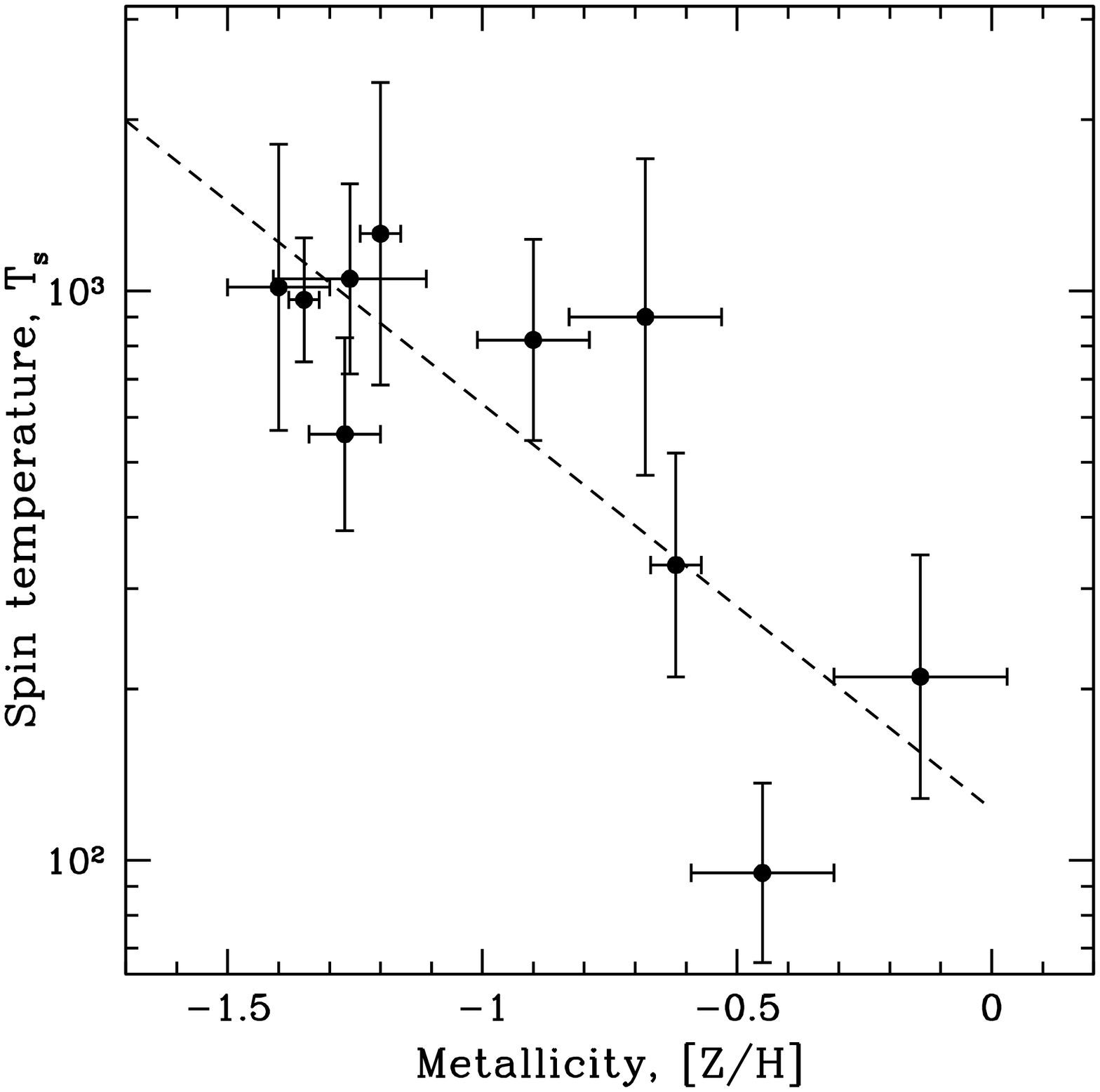,height=3.5truein,width=3.5truein}
\caption{The spin temperature $\ts$ plotted against metallicity, [Z/H], for 
[A]~(left panel) the 26~DLAs of the full sample, and [B]~(right panel) the ten DLAs 
with measurements of both $\ts$ and [Z/H]. The dashed line shows the linear fit 
to the relation between [Z/H] and $\ts$.}
\label{fig:tszn}
\end{figure*}


\section{Results: An anti-correlation between $\ts$ and [Z/H]}
\label{sec:results}

Fig.~\ref{fig:tszn}[A] shows $\ts$ plotted versus [Z/H] for the 26~DLAs of the full sample; 
it is apparent that low $\ts$ values ($\lesssim 300$~K) are obtained at high metallicities 
([Z/H]$\gtrsim -0.6$), while high $\ts$ values $\gtrsim 700$~K are found at low metallicities,
[Z/H]$\lesssim -1$. We used the non-parametric generalized Kendall rank correlation 
statistic \citep{brown74,isobe86}, as implemented in the {\sc ASURV} package (the BHK 
statistic), to test for a correlation between $\ts$ and [Z/H], treating the latter as the 
independent variable. This allows limits in both variables to be treated consistently. 
Errors on individual measurements were handled through a Monte-Carlo approach, using the 
measured values and $1\sigma$ errors for each absorber to generate $10^4$ sets of 26 
pairs of [Z/H] and $\ts$ values. The statistical significance of the BHK statistic 
quoted below is the average of values obtained in these $10^4$ trials. For the 
full sample, the anti-correlation is detected at $\sim 3.6 \sigma$ significance; 
the probability of this arising by chance is $\sim 3 \times 10^{-4}$. If we exclude
the $z = 0.524$ DLA towards 0235+164 (whose metallicity is from an X-ray study), 
and use [Fe/H] as a lower limit to [Z/H] for the DLA towards 0827+243, 
the anti-correlation is detected at $\sim 3.1 \sigma$ significance, showing that
it does not arise from incorrect metallicity estimates in these absorbers.

A linear regression analysis was used to obtain the best-fit relation between
$\ts$ and [Z/H], using the ten systems with measurements of both $\ts$ and 
[Z/H] (i.e. excluding all limits), all of which also have covering 
factor estimates; these are shown in Fig.~\ref{fig:tszn}[B]. The BCES(Y/X) estimator 
\citep{akritas96} was used for this purpose, treating [Z/H] as the independent variable X; 
this method takes into account measurement errors on both variables, as well as the 
possibility that these errors are correlated (which applies here, as $\ts$ and [Z/H] 
are both derived from $\nhi$). We obtain $\ts = (-0.68 \pm 0.17) \times {\rm [Z/H]} + 
(2.13 \pm 0.21)$; the fit is shown as a dashed line in Figs.~\ref{fig:tszn}[A] and [B].
Excluding the DLAs towards 0235+164 and 0827+243 yields $\ts = (-0.77 \pm 0.40) \times 
{\rm [Z/H]} + (2.02 \pm 0.48)$, from eight DLAs with measurements of $\ts$ and [Z/H].

We also tested for a correlation between the dust depletion factor [Z/Fe] and 
$\ts$, using the 21~systems with estimates of both quantities. The anti-correlation 
between [Z/Fe] and $\ts$ has $\sim 2.2\sigma$ significance, reducing to 
$\sim 1.6\sigma$ significance on excluding the possibly-unreliable X-ray 
metallicity estimate in the DLA towards 0235+164. We thus do not find 
significant evidence for a relation between [Z/Fe] and $\ts$, although 
this sub-sample contains few DLAs with low spin temperatures. 

\citet{curran07c} note that an observed anti-correlation between $\ts$ 
and metallicity might arise due to unknown low covering factors.
We hence estimated the BHK statistic for the sub-sample of 20~DLAs with covering 
factor estimates \citep{kanekar09a}. The anti-correlation between $\ts$ and [Z/H] 
is then detected at $\sim 3\sigma$ significance, with a probability of $\sim 0.003$ 
of chance occurrence. Low covering factors are thus unlikely to be the cause of the 
anti-correlation between $\ts$ and [Z/H]. 


It is possible that the observed anti-correlation between $\ts$ and [Z/H] is not a 
``primary'' relation, but stems from an underlying relation between other 
physical quantities (e.g. velocity spread and metallicity; \citealp{ledoux06}). 
For example, \citet{curran07c} argue that an observed correlation between $\ts$ 
and [Z/H] might arise due to a relation between the velocity spread of 
\hi~21cm absorption $\Delta V_{\rm 21}$ and the rest equivalent width of 
the \mgtwo\ line, $\wmgtwo$, each tracing the dynamics of the absorber.
They note that $\ts \propto {\rm N_{HI}}/\left[\int \tau_{\rm 21} {\rm d}V \right] \approx 
{\rm N_{HI}} / \left[ \tau_{\rm 21,max} \Delta V_{\rm 21} \right]$,
suggesting that $\ts \propto 1/\Delta V_{\rm 21}$. $\wmgtwo$ is known to correlate with 
metallicity \citep{murphy07b}, so, if $\Delta V_{\rm 21}$ also correlates with $\wmgtwo$, 
the three relations could yield an observed anti-correlation between $\ts$ and [Z/H].
However, the \hi\ column density in DLAs does not correlate with either [Z/H] or 
$\wmgtwo$ (e.g. \citealt{rao06}), and the DLAs of Table~\ref{tab:met} have 
$\nhi$ values extending over $\approx 1.5$~dex. If the $\ts$--[Z/H] relation arises 
due to an underlying relation between $\Delta V_{\rm 21}$ and $\wmgtwo$, this spread 
in $\nhi$ values would imply that the $\ts$--[Z/H] relation would be {\it weaker} than 
the relation between $\Delta V_{\rm 21}$ and $\wmgtwo$ (and vice-versa, if the $\ts$--[Z/H] 
relation is the ``primary'' relation). In other words, the ``primary'' relation 
would have the highest statistical significance in samples of similar size, as 
uncorrelated (and variable) parameters like $\nhi$ would dilute the significance 
of any derived relations.

\citet{kanekar09b} used the BHK test to test for correlations between 
$\wmgtwo$ and both the \hi~21cm velocity spread $\Delta V_{\rm 21}$ and 
the integrated \hi~21cm optical depth $\int \tau_{\rm 21} {\rm d}V$, in a sample 
of Mg{\sc ii} absorbers and DLAs. They found the putative correlation between 
$\Delta V_{\rm 21}$ and $\wmgtwo$ to have $\sim 1.7\sigma$ significance (with 
23~absorbers), and that between $\int \tau_{\rm 21} {\rm d}V$ and $\wmgtwo$ 
to have $\sim 2.2\sigma$ significance (with 24~absorbers). Both these trends 
are significantly weaker than the relation between $\ts$ and [Z/H] found 
here ($\sim 3.6\sigma$ significance), in samples of similar size. It is thus 
unlikely that the anti-correlation between $\ts$ and [Z/H] arises from an 
underlying relation between $\wmgtwo$ and either $\Delta V_{\rm 21}$ or 
$\int \tau_{\rm 21} {\rm d}V$. The present data indicate that the 
anti-correlation between $\ts$ and [Z/H] is the ``primary'' relation, given 
its higher statistical significance. 

\section{Discussion}

High spin temperature estimates in high-$z$ DLAs have been the focus of much debate, 
with suggestions that these might arise due to different sightlines at optical and radio 
wavelengths \citep{wolfe03b} or low covering factors \citep{curran05}. It is possible that 
individual $\ts$ measurements might be in error due to these effects. However, recent
low-frequency VLBI studies have shown that high $\ts$ values are not the result
of low DLA covering factors \citep{kanekar09a}. Further, the anti-correlation 
between $\ts$ and [Z/H] is detected in the sub-sample of DLAs with covering factor 
measurements, indicating that this relation too is not caused by low covering factors. 
Finally, differing optical and radio sightlines (e.g.  if the radio emission is extended 
on scales larger than the size of the optical quasar) would result in $\ts$ estimates 
that are uncorrelated with the ``true'' $\ts$ values along the optical sightline. 
This would {\it weaken} any underlying correlation between $\ts$ and a quantity like 
[Z/H], measured along the pencil beam towards the optical quasar.  The detection of 
the predicted anti-correlation between $\ts$ and [Z/H] thus indicates that line-of-sight 
issues are also not the source of the observed high $\ts$ values. We conclude that the 
high $\ts$ estimates in high-$z$ DLAs are most likely to arise due to larger WNM 
fractions in DLAs than typical in local spiral galaxies.

In the local universe, the metallicity and mass of galaxies are known to be correlated
(e.g. \citealp{tremonti04}).  A similar correlation, between metallicity and stellar mass,
has been found in emission-selected, high-$z$ galaxies (e.g. \citealp{savaglio05}). 
Evidence for a mass-metallicity relation in DLAs is unclear. \citet{ledoux06} argue 
that the observed correlation between the kinematic width of unsaturated low-ionization 
metal profiles $\Delta V_{90}$ and metallicity in high-$z$ DLAs arises from an 
underlying mass-metallicity relation. Conversely, \citet{zwaan08} note that 
$\Delta V_{90}$ is a weak indicator of galaxy mass at $z = 0$, as the kinematic 
width measured along a pencil beam depends on the galaxy inclination and the impact parameter. 
However, the above correlation between $\Delta V_{90}$ and metallicity has been reproduced in 
simulations \citep{pontzen08}, which suggest that it does stem from a mass-metallicity 
relation. \citet{pontzen08} find that sightlines through DLAs with low virial masses, 
$M_{\rm vir} < 10^{9.5} \: M_\odot$, typically yield low metallicities, [Z/H]$\lesssim -1.7$, 
while the typical metallicities are [Z/H]$\sim -1.2$ in intermediate-mass galaxies 
($M_{\rm vir} \sim 10^{9.5 - 10.5} \: M_\odot$), and [Z/H]$\gtrsim -1$ in high-mass 
systems ($M_{\rm vir} > 10^{10.5} \: M_\odot$; for comparison, $M_{\rm vir} \sim 10^{12} \: 
M_\odot$ for the Milky Way; e.g. \citealp{xue08}). Interestingly, \citet{prochaska08} 
find a tight correlation between the rest equivalent width in the saturated 
Si{\sc ii}$\lambda$1526 line $W_{\rm 1526}$ and metallicity, and argue that 
$W_{\rm 1526}$ tracks dynamical motions in the halos of DLAs, with the 
correlation between $W_{1526}$ and [Z/H] arising due to a mass-metallicity relation 
of the same slope as that seen in low-$z$ galaxies.

If a mass-metallicity relation is present in DLAs, low-mass DLAs would have low 
metallicities. Low-mass galaxies also have low thermal pressures, while a high pressure is 
needed for the formation of a stable multi-phase medium at a low metallicity. Fig.~6 of 
\citet{wolfire95} shows that, at low pressures and metallicities, \hi\ exists mostly
in the warm phase. Sightlines through such galaxies would thus typically yield a high 
spin temperature. Conversely, the mass-metallicity relation implies that high-mass DLAs 
would have high metallicities, along with high thermal pressures. Such galaxies would 
have significant CNM fractions, with most sightlines yielding low spin temperatures. 
These effects would yield the anti-correlation between $\ts$ and [Z/H] detected here, 
due to the paucity of cooling routes in low-metallicity galaxies 
\citep{norman97,kanekar01a,kanekar04}. The anti-correlation between $\ts$ and 
[Z/H] is thus consistent with the presence of a mass-metallicity relation in DLAs 
\citep{ledoux06,prochaska08}; low-metallicity DLAs are likely to have high $\ts$ 
values due to their low CNM fractions.

While the relation between $\ts$ and [Z/H] is consistent with the presence of a 
mass-metallicity relation in DLAs, it is also possible that the $\ts$--[Z/H] 
relation is a local one, arising due to line-of-sight issues. For disk galaxies, 
the cross-section for DLA incidence is largest for sightlines through the outskirts 
of the galaxy. If high-$z$ DLAs have steep metallicity gradients, such sightlines 
would typically encounter low metallicities (e.g. \citealt{zwaan05}). The lack of 
{\it local} cooling routes could then result in low CNM fractions, and high $\ts$ 
values along these sightlines. Conversely, sightlines through the central regions 
of large disk galaxies would typically have high metallicities and high CNM fractions. 
Such line-of-sight effects could thus also yield the anti-correlation of 
Fig.~\ref{fig:tszn}. While such steep metallicity gradients have not been seen in 
local galaxies and low-$z$ DLAs (typical gradients are $\sim -0.03 - -0.04$~dex~kpc$^{-1}$; 
\citealt{chen05,bresolin09}), they are not ruled out in high-$z$ DLAs. We note,
in passing, that the $z \sim 0.524$ DLA towards 0827+243 has one of the largest DLA 
impact parameters ($\sim 27$~kpc; \citealt{steidel02}), and yet has both a low $\ts$ 
($330$~K) and a high metallicity, [Z/H]$=-0.6$.

Fig.~\ref{fig:tszn}[A] shows that low $\ts$ values are only found in DLAs with 
[Z/H]$\gtrsim -0.6$. Only 13~of the 153~DLAs at $z > 1.6$ in the sample
of \citet{prochaska07} have such high metallicities; the majority of high-$z$ DLAs
have [Z/H]$< -1.0$. The anti-correlation between $\ts$ and [Z/H] then implies that 
most high~$z$ DLAs have high $\ts$ values. Conversely, six of seventeen DLAs at 
$z \lesssim 1$ have [Z/H]$ \ge -0.6$ (e.g. \citealp{prochaska03a,kulkarni05}).
While the number of metallicity measurements at $z<1$ is yet small, the present 
data indicate a higher fraction of high-metallicity DLAs at low redshifts. If 
so, the $\ts$--[Z/H] anti-correlation implies that the fraction of DLAs with 
high CNM fraction should also be larger at $z \lesssim 1$. This is consistent 
with the observed increase in the detection rate of \hi~21cm absorption in 
DLAs at $z \lesssim 1$ \citep{kanekar09b}.

Finally, \citet{wolfe03b} found that the strong CII* lines detected in half their 
DLAs could not be explained by absorption in pure WNM, and implied the presence 
of cold \hi. They argue that \hi\ in DLAs without CII* absorption is likely to be 
predominantly warm, but that a two-phase model with a CNM fraction of $\sim 50$\% 
is consistent with the observed bolometric luminosity in systems with CII* detections. 
However, lower CNM fractions are not ruled out. For example, the $z \sim 3.39$ DLA 
towards PKS~0201+113 shows strong CII* absorption \citep{wolfe03b}, but has a low 
metallicity, [S/H]$= -1.26$ \citep{ellison01b}, and a low CNM fraction, 
$\lesssim 17$\% \citep{kanekar07}. This suggests that, while DLAs showing CII* 
absorption contain {\it some} CNM, the \hi\ content of low-metallicity absorbers 
is still likely to be dominated by the WNM (see also \citealp{srianand05}).

\section{Summary}
\label{sec:summary}

We report the detection of an anti-correlation between spin temperature 
and metallicity (with $\sim 3.6 \sigma$ significance in the non-parametric BHK test) 
in a sample of 26~DLAs at $0.09 < z < 3.45$. For 20~systems, the absorber covering 
factor has been estimated from low-frequency VLBI studies; the anti-correlation between $\ts$ 
and [Z/H] is detected here at $\sim 3\sigma$ significance. A linear regression
analysis using the BCES estimator finds the relation $\ts = (-0.68 \pm 0.17) \times {\rm [Z/H]} 
+ (2.13 \pm 0.21)$ between $\ts$ and [Z/H]. Low spin temperatures, $\ts \lesssim 300$~K, 
are only found in high-metallicity DLAs (with [Z/H]$\gtrsim -0.6$), while high 
$\ts$ values ($\gtrsim 700$~K) are obtained in low-metallicity ([Z/H]$< -1$) absorbers. 
The fact that a relation is seen between $\ts$ and [Z/H] implies that the high 
$\ts$ values in DLAs are unlikely to be an artifact arising from differences between 
radio and optical sightlines through the absorbers, and are instead likely to reflect 
the underlying gas temperature distribution. The majority of DLAs at $z > 1.6$ have 
[Z/H]$< -1$, implying that most of the \hi\ in DLAs must be in the warm phase, 
with small CNM fractions. The observed anti-correlation between $\ts$ and [Z/H] is 
consistent with independent evidence for the presence of a mass-metallicity relation 
in DLAs. The majority of high-$z$ DLAs are likely to be galaxies of low mass and 
metallicity, with most of the neutral gas in the warm phase.

\acknowledgments 
We thank Sara Ellison and an anonymous referee for comments on an earlier draft.
Part of this work was carried out during a visit by NK to ESO, Santiago, under an
ESO Visiting Fellowship; he thanks ESO for support and hospitality, and is also grateful 
for support from the Max-Planck Society and the Alexander von Humboldt Foundation.

\bibliographystyle{apj}


\begin{thebibliography}{50}
\expandafter\ifx\csname natexlab\endcsname\relax\def\natexlab#1{#1}\fi

\bibitem[{{Akerman} {et~al.}(2005){Akerman}, {Ellison}, {Pettini}, \&
  {Steidel}}]{akerman05}
{Akerman}, C.~J., {Ellison}, S.~L., {Pettini}, M., \& {Steidel}, C.~C. 2005,
  A\&A, 440, 499

\bibitem[{{Akritas} \& {Bershady}(1996)}]{akritas96}
{Akritas}, M.~G. \& {Bershady}, M.~A. 1996, ApJ, 470, 706

\bibitem[{{Boisse} {et~al.}(1998){Boisse}, {Le Brun}, {Bergeron}, \&
  {Deharveng}}]{boisse98}
{Boisse}, P., {Le Brun}, V., {Bergeron}, J., \& {Deharveng}, J.-M. 1998, A\&A,
  333, 841

\bibitem[{Braun \& Walterbos(1992)}]{braun92}
Braun, R. \& Walterbos, R. 1992, ApJ, 386, 120

\bibitem[{{Bresolin} {et~al.}(2009){Bresolin}, {Ryan-Weber}, {Kennicutt}, \&
  {Goddard}}]{bresolin09}
{Bresolin}, F., {Ryan-Weber}, E., {Kennicutt}, R.~C., \& {Goddard}, Q. 2009,
  ApJ, 695, 580

\bibitem[{Briggs {et~al.}(2001)Briggs, {de Bruyn}, \& Vermeulen}]{briggs01}
Briggs, F.~H., {de Bruyn}, A.~G., \& Vermeulen, R.~C. 2001, A\&A, 373, 113

\bibitem[{{Brown} {et~al.}(1974){Brown}, {Hollander}, \& {Korwar}}]{brown74}
{Brown}, B.~W.~M., {Hollander}, M., \& {Korwar}, R.~M. 1974, in Reliability and
  Biometry, ed. F.~Proschan \& R.~J. Serfling (Philadelphia: SIAM), 327

\bibitem[{Carilli {et~al.}(1996)Carilli, Lane, {de Bruyn}, Braun, \&
  Miley}]{carilli96}
Carilli, C.~L., Lane, W.~M., {de Bruyn}, A.~G., Braun, R., \& Miley, G.~K.
  1996, AJ, 111, 1830

\bibitem[{{Chen} {et~al.}(2005){Chen}, {Kennicutt}, \& {Rauch}}]{chen05}
{Chen}, H.-W., {Kennicutt}, Jr., R.~C., \& {Rauch}, M. 2005, ApJ, 620, 703

\bibitem[{Chengalur \& Kanekar(2000)}]{chengalur00}
Chengalur, J.~N. \& Kanekar, N. 2000, MNRAS, 318, 303

\bibitem[{{Curran} {et~al.}(2005){Curran}, {Murphy}, {Pihlstr{\"o}m}, {Webb},
  \& {Purcell}}]{curran05}
{Curran}, S.~J., {Murphy}, M.~T., {Pihlstr{\"o}m}, Y.~M., {Webb}, J.~K., \&
  {Purcell}, C.~R. 2005, MNRAS, 356, 1509

\bibitem[{{Curran} {et~al.}(2007{\natexlab{a}}){Curran}, {Tzanavaris},
  {Murphy}, {Webb}, \& {Pihlstroem}}]{curran07a}
{Curran}, S.~J., {Tzanavaris}, P., {Murphy}, M.~T., {Webb}, J.~K., \&
  {Pihlstroem}, Y.~M. 2007{\natexlab{a}}, MNRAS, 381, L6

\bibitem[{{Curran} {et~al.}(2007{\natexlab{b}}){Curran}, {Tzanavaris},
  {Pihlstr{\"o}m}, \& {Webb}}]{curran07c}
{Curran}, S.~J., {Tzanavaris}, P., {Pihlstr{\"o}m}, Y.~M., \& {Webb}, J.~K.
  2007{\natexlab{b}}, MNRAS, 382, 1331

\bibitem[{{Ellison} {et~al.}(2001){Ellison}, {Pettini}, {Steidel}, \&
  {Shapley}}]{ellison01b}
{Ellison}, S.~L., {Pettini}, M., {Steidel}, C.~C., \& {Shapley}, A.~E. 2001,
  ApJ, 549, 770

\bibitem[{{Ellison} {et~al.}(2008){Ellison}, {York}, {Pettini}, \&
  {Kanekar}}]{ellison08}
{Ellison}, S.~L., {York}, B.~A., {Pettini}, M., \& {Kanekar}, N. 2008, MNRAS,
  388, 1349

\bibitem[{{Isobe} {et~al.}(1986){Isobe}, {Feigelson}, \& {Nelson}}]{isobe86}
{Isobe}, T., {Feigelson}, E.~D., \& {Nelson}, P.~I. 1986, ApJ, 306, 490

\bibitem[{{Junkkarinen} {et~al.}(2004){Junkkarinen}, {Cohen}, {Beaver},
  {Burbidge}, {Lyons}, \& {Madejski}}]{junkkarinen04}
{Junkkarinen}, V.~T., {Cohen}, R.~D., {Beaver}, E.~A., {Burbidge}, E.~M.,
  {Lyons}, R.~W., \& {Madejski}, G. 2004, ApJ, 614, 658

\bibitem[{Kanekar \& Briggs(2004)}]{kanekar04}
Kanekar, N. \& Briggs, F.~H. 2004, New Astr. Rev., 48, 1259

\bibitem[{Kanekar \& Chengalur(2001)}]{kanekar01a}
Kanekar, N. \& Chengalur, J.~N. 2001, A\&A, 369, 42

\bibitem[{Kanekar \& Chengalur(2003)}]{kanekar03}
---. 2003, A\&A, 399, 857

\bibitem[{{Kanekar} \& {Chengalur}(2005)}]{kanekar05b}
{Kanekar}, N. \& {Chengalur}, J.~N. 2005, in IAU Colloq. 199: Probing Galaxies
  through Quasar Absorption Lines, ed. P.~{Williams}, C.-G. {Shu}, \&
  B.~{Menard} (Cambridge University Press, Cambridge), 156

\bibitem[{Kanekar {et~al.}(2007)Kanekar, Chengalur, \& Lane}]{kanekar07}
Kanekar, N., Chengalur, J.~N., \& Lane, W.~M. 2007, MNRAS, 375, 1528

\bibitem[{{Kanekar} {et~al.}(2009{\natexlab{a}}){Kanekar}, {Lane}, {Momjian},
  {Briggs}, \& {Chengalur}}]{kanekar09a}
{Kanekar}, N., {Lane}, W.~M., {Momjian}, E., {Briggs}, F.~H., \& {Chengalur},
  J.~N. 2009{\natexlab{a}}, MNRAS, 394, L61

\bibitem[{{Kanekar} {et~al.}(2009{\natexlab{b}}){Kanekar}, {Prochaska},
  {Ellison}, \& {Chengalur}}]{kanekar09b}
{Kanekar}, N., {Prochaska}, J.~X., {Ellison}, S.~L., \& {Chengalur}, J.~N.
  2009{\natexlab{b}}, MNRAS, 396, 385

\bibitem[{Kanekar {et~al.}(2006)Kanekar, Subrahmanyan, Ellison, Lane, \&
  Chengalur}]{kanekar06}
Kanekar, N., Subrahmanyan, R., Ellison, S.~L., Lane, W.~M., \& Chengalur, J.~N.
  2006, MNRAS, 370, L46

\bibitem[{{Kulkarni} {et~al.}(2005){Kulkarni}, {Fall}, {Lauroesch}, {York},
  {Welty}, {Khare}, \& {Truran}}]{kulkarni05}
{Kulkarni}, V.~P., {Fall}, S.~M., {Lauroesch}, J.~T., {York}, D.~G., {Welty},
  D.~E., {Khare}, P., \& {Truran}, J.~W. 2005, ApJ, 618, 68

\bibitem[{{Ledoux} {et~al.}(2006){Ledoux}, {Petitjean}, {Fynbo}, {M{\o}ller},
  \& {Srianand}}]{ledoux06}
{Ledoux}, C., {Petitjean}, P., {Fynbo}, J.~P.~U., {M{\o}ller}, P., \&
  {Srianand}, R. 2006, A\&A, 457, 71

\bibitem[{{Lodders}(2003)}]{lodders03}
{Lodders}, K. 2003, ApJ, 591, 1220

\bibitem[{{Murphy} {et~al.}(2007){Murphy}, {Curran}, {Webb}, {M{\'e}nager}, \&
  {Zych}}]{murphy07b}
{Murphy}, M.~T., {Curran}, S.~J., {Webb}, J.~K., {M{\'e}nager}, H., \& {Zych},
  B.~J. 2007, MNRAS, 376, 673

\bibitem[{{Nestor} {et~al.}(2008){Nestor}, {Pettini}, {Hewett}, {Rao}, \&
  {Wild}}]{nestor08}
{Nestor}, D., {Pettini}, M., {Hewett}, P., {Rao}, S., \& {Wild}, V. 2008, MNRAS
  (accepted; arxiv/0808.2470)

\bibitem[{{Norman} \& {Spaans}(1997)}]{norman97}
{Norman}, C.~A. \& {Spaans}, M. 1997, ApJ, 480, 145

\bibitem[{{Pontzen} {et~al.}(2008){Pontzen}, {Governato}, {Pettini}, {Booth},
  {Stinson}, {Wadsley}, {Brooks}, {Quinn}, \& {Haehnelt}}]{pontzen08}
{Pontzen}, A., {Governato}, F., {Pettini}, M., {Booth}, C.~M., {Stinson}, G.,
  {Wadsley}, J., {Brooks}, A., {Quinn}, T., \& {Haehnelt}, M. 2008, MNRAS, 390,
  1349

\bibitem[{{Prochaska} {et~al.}(2008){Prochaska}, {Chen}, {Wolfe},
  {Dessauges-Zavadsky}, \& {Bloom}}]{prochaska08}
{Prochaska}, J.~X., {Chen}, H.-W., {Wolfe}, A.~M., {Dessauges-Zavadsky}, M., \&
  {Bloom}, J.~S. 2008, ApJ, 672, 59

\bibitem[{{Prochaska} {et~al.}(2003{\natexlab{a}}){Prochaska}, {Gawiser},
  {Wolfe}, {Castro}, \& {Djorgovski}}]{prochaska03a}
{Prochaska}, J.~X., {Gawiser}, E., {Wolfe}, A.~M., {Castro}, S., \&
  {Djorgovski}, S.~G. 2003{\natexlab{a}}, ApJ, 595, L9

\bibitem[{{Prochaska} {et~al.}(2003{\natexlab{b}}){Prochaska}, {Gawiser},
  {Wolfe}, {Cooke}, \& {Gelino}}]{prochaska03b}
{Prochaska}, J.~X., {Gawiser}, E., {Wolfe}, A.~M., {Cooke}, J., \& {Gelino}, D.
  2003{\natexlab{b}}, ApJS, 147, 227

\bibitem[{Prochaska {et~al.}(2007)Prochaska, Wolfe, Howk, Gawiser, Burles, \&
  Cooke}]{prochaska07}
Prochaska, J.~X., Wolfe, A.~M., Howk, J.~C., Gawiser, E., Burles, S.~M., \&
  Cooke, J. 2007, ApJS, 171, 29

\bibitem[{{Rao} {et~al.}(2006){Rao}, {Turnshek}, \& {Nestor}}]{rao06}
{Rao}, S.~M., {Turnshek}, D.~A., \& {Nestor}, D.~B. 2006, ApJ, 636, 610

\bibitem[{{Savaglio} {et~al.}(2005){Savaglio}, {Glazebrook}, {Le Borgne},
  {Juneau}, {Abraham}, {Chen}, {Crampton}, {McCarthy}, {Carlberg}, {Marzke},
  {Roth}, {J{\o}rgensen}, \& {Murowinski}}]{savaglio05}
{Savaglio}, S. et al. 2005, ApJ, 635, 260

\bibitem[{{Srianand} {et~al.}(2005){Srianand}, {Petitjean}, {Ledoux},
  {Ferland}, \& {Shaw}}]{srianand05}
{Srianand}, R., {Petitjean}, P., {Ledoux}, C., {Ferland}, G., \& {Shaw}, G.
  2005, MNRAS, 362, 549

\bibitem[{{Steidel} {et~al.}(2002){Steidel}, {Kollmeier}, {Shapley},
  {Churchill}, {Dickinson}, \& {Pettini}}]{steidel02}
{Steidel}, C.~C., {Kollmeier}, J.~A., {Shapley}, A.~E., {Churchill}, C.~W.,
  {Dickinson}, M., \& {Pettini}, M. 2002, ApJ, 570, 526

\bibitem[{{Tremonti} {et~al.}(2004){Tremonti}, {Heckman}, {Kauffmann},
  {Brinchmann}, {Charlot}, {White}, {Seibert}, {Peng}, {Schlegel}, {Uomoto},
  {Fukugita}, \& {Brinkmann}}]{tremonti04}
{Tremonti}, C.~A. et al. 2004, ApJ, 613, 898

\bibitem[{Wolfe \& Davis(1979)}]{wolfe79}
Wolfe, A.~M. \& Davis, M.~M. 1979, AJ, 84, 699

\bibitem[{Wolfe {et~al.}(2003)Wolfe, Gawiser, \& Prochaska}]{wolfe03b}
Wolfe, A.~M., Gawiser, E., \& Prochaska, J.~X. 2003, ApJ, 593, 235

\bibitem[{Wolfe {et~al.}(2005)Wolfe, Gawiser, \& Prochaska}]{wolfe05}
---. 2005, ARA\&A, 43, 861

\bibitem[{{Wolfe} {et~al.}(2008){Wolfe}, {Jorgenson}, {Robishaw}, {Heiles}, \&
  {Prochaska}}]{wolfe08}
{Wolfe}, A.~M., {Jorgenson}, R.~A., {Robishaw}, T., {Heiles}, C., \&
  {Prochaska}, J.~X. 2008, Nature, 455, 638

\bibitem[{{Wolfire} {et~al.}(1995){Wolfire}, {Hollenbach}, {McKee}, {Tielens},
  \& {Bakes}}]{wolfire95}
{Wolfire}, M.~G., {Hollenbach}, D., {McKee}, C.~F., {Tielens}, A.~G.~G.~M., \&
  {Bakes}, E.~L.~O. 1995, ApJ, 443, 152

\bibitem[{{Xue} {et~al.}(2008){Xue}, {Rix}, {Zhao}, {Re Fiorentin}, {Naab},
  {Steinmetz}, {van den Bosch}, {Beers}, {Lee}, {Bell}, {Rockosi}, {Yanny},
  {Newberg}, {Wilhelm}, {Kang}, {Smith}, \& {Schneider}}]{xue08}
{Xue}, X.~X. et al. 2008, ApJ, 684, 1143

\bibitem[{{York} {et~al.}(2007){York}, {Kanekar}, {Ellison}, \&
  {Pettini}}]{york07}
{York}, B.~A., {Kanekar}, N., {Ellison}, S.~L., \& {Pettini}, M. 2007, MNRAS,
  382, L53

\bibitem[{{Zwaan} {et~al.}(2008){Zwaan}, {Walter}, {Ryan-Weber}, {Brinks}, {de
  Blok}, \& {Kennicutt}}]{zwaan08}
{Zwaan}, M., {Walter}, F., {Ryan-Weber}, E., {Brinks}, E., {de Blok}, W.~J.~G.,
  \& {Kennicutt}, R.~C. 2008, AJ, 136, 2886

\bibitem[{{Zwaan} {et~al.}(2005){Zwaan}, {van der Hulst}, {Briggs},
  {Verheijen}, \& {Ryan-Weber}}]{zwaan05}
{Zwaan}, M.~A., {van der Hulst}, J.~M., {Briggs}, F.~H., {Verheijen}, M.~A.~W.,
  \& {Ryan-Weber}, E.~V. 2005, MNRAS, 364, 1467

\end{thebibliography}

\end{document}